\documentclass{elsart}
\usepackage{graphicx}
\usepackage{dcolumn}
\usepackage{bm}
\usepackage{amsmath}
\usepackage{amsfonts}
\usepackage{amssymb}
\usepackage{subfigure}
\providecommand{\U}[1]{\protect\rule{.1in}{.1in}}

\begin{document}
\begin{frontmatter}
\title{Relation between the usual and the entanglement temperature, in a simple quantum system}
\author{Alejandro Romanelli},
\author{Raul Donangelo},
\author{Andr\'es Vallejo},
\address{Instituto de F\'{\i}sica, Facultad de Ingenier\'{\i}a\\
Universidad de la Rep\'ublica\\ C.C. 30, C.P. 11000, Montevideo, Uruguay}


\begin{abstract}
We develop a thermodynamical theory to describe the behavior of the entanglement between a single two-level atom with a
single mode of the electromagnetic field. The resonant Jaynes-Cummings model is used to study both the entanglement
thermodynamics, in particular the entanglement temperature, and its connection with the average number of photons in the
optical cavity. We find that this entanglement temperature has a strong dependence with the  initial conditions of the
atom. We show that the entanglement temperature between the photons and the atom defined in this work is the same
temperature obtained within the Jaynes-Cummings model at finite temperature developed in the Thermo-Field Dynamics
formalism.
\end{abstract}
\end{frontmatter}

\section{Introduction}
Concepts such as thermodynamic equilibrium seem impossible to reconcile with the idea of isolated quantum systems since
such systems follow unitary evolutions and do not reach a final stationary equilibrium state. Of course, a completely
isolated quantum system is an idealization, constructed as a help to understand some phenomena displayed by real systems
which may be regarded as approximately isolated. However, we recently \cite{alejo2012,gustavo,renato2014,alejo2015}
introduced the concept of temperature for an isolated quantum system which evolves in a composite Hilbert space. To do
this we consider the quantum walk on the line (QW) (see \cite{QW} and references therein). The QW is a natural
generalization of the classical random walk in the frame of quantum computation and quantum information processing and it
receiving much attention recently \cite{childs,Linden,Alejo3,german2013}.

In our above mentioned works, we have developed a thermodynamic theory to describe the behavior of the entanglement
between the coin and position degrees of freedom of the QW. Henceforth, we call ``entanglement temperature'' to the
temperature associated to the entanglement between different degrees of freedom of an isolated quantum system. We have
shown that, in spite of the evolution being unitary, in the QW a steady state is established after a Markovian transient
stage. Those studies suggest that, if a quantum dynamics develops in a composite Hilbert space (\emph{i.e.} the tensor
product of several sub-spaces), then the behavior of an operator that belongs only to one of the sub-spaces may
camouflage the unitary character of the global evolution. However it is not clear what is the relation between the usual
temperature and the entanglement temperature. This question cannot be answered using the QW because it is an abstract
mathematical model; to do this we need a real physical model where both temperatures, the usual and the entanglement
temperature, emerge naturally. In order to answer it, we have here chosen one of the simplest and most interesting
quantum models, the one known as the Jaynes-Cummings model (JCM) \cite{Jaynes1,Jaynes2}, that studies the interaction
between radiation and matter.

The JCM considers the interaction between a single two-level atom with a single mode of the electromagnetic field. The
coupling between the atom and the field is characterized by a Rabi frequency, and a loss of excitation in the atom
appears as a gain in excitation of the field. The collapse and the eventual revival of the Rabi oscillation, described by
the analytical solution of the JCM, is a direct evidence of the quantum nature of radiation. The use of the JCM has
permitted to elucidate basic properties of quantum entanglement as well as some aspects of the relationship between
classical and quantum physics. Since it was proposed, the phenomenon has been of permanent interest in the quantum theory
of interactions. About $30$ years ago it was found that the model exhibits highly non-classic behavior, and the
possibility of experimental realizations appeared. The relative simplicity of the JCM and its extensions has drawn much
attention in the physics community and, more recently, in the field of the quantum computing \cite{Chuang,alejoJaynes}.

Also in the $80$'s, the Thermo Field Dynamics (TFD) formalism \cite{Thermo,Thermo1} was applied to the JCM. The TFD is a
method, developed in the $70$'s by Takahashi and Umezawa \cite{Thermo0}, for describing Quantum Mechanical systems at
finite temperature. Using this method, it is possible to describe the statistical average of an observable at finite
temperature as a pure state expectation value. Thus, within the TFD formalism, one does not need to deal with a mixed
state, which is a statistical ensemble of pure states at finite temperature. In return for the above advantage, the TFD
introduces the so-called tilde particles corresponding to ordinary particles, thus doubling the dimension of the Hilbert
space associated to the system. In the TFD method the ordinary particles and the introduced tilde particles represent the
dynamical degrees of freedom and the thermal degrees of freedom, respectively.

In the present work we connect, within the JCM, the TFD thermodynamics with the entanglement thermodynamics presented in
our previous works \cite{alejo2012,gustavo,renato2014}. The paper is organized as follows. In the next section we review
the usual JCM and study the photon thermodynamics in that model. In third section we develop the entanglement
thermodynamics for the JCM and study its connection to the TFD-defined temperature. Finally, in the last section we draw
some conclusions.
\section{Jaynes-Cummings model}
We consider the ordinary JCM \cite{Jaynes2}, composed by a single two-state atom in an optical cavity, interacting with a
single quantized mode, with frequency $\omega$. The Hilbert space of the JCM has the form of a tensor product
\begin{equation}
\mathcal{H}=\mathcal{H}_{\mathrm{N}}\otimes\mathcal{H}_{\mathrm{A}}, \label{espacio}
\end{equation}
\noindent where the photon space, $\mathcal{H}_{\mathrm{N}}$, is spanned by the unitary orthonormal vectors of the photon
number state $\left\{ \left\vert n\right\rangle \right\} $, and the atom space, $\mathcal{H}_{\mathrm{A}}$, is spanned by
the two orthonormal quantum states $\left\{ \left\vert e\right\rangle, \left\vert f\right\rangle \right\}$ that represent
the excited and fundamental states of the atom, respectively. Note that the set $\left\{|n, e\rangle,|n,
f\rangle\right\}$, where $|n,e\rangle=|n\rangle|e\rangle$ and $ |n, f\rangle=|n\rangle|f\rangle$, is an orthonormal base
in the JCM Hilbert space.

In this model, if the atom excitation frequency $\omega _{a}$ is close to $\omega $, then the system is near the
resonance and it is possible to use the rotating wave approximation. In this case, and removing the field vacuum energy,
the system Hamiltonian is
\begin{equation}
H=\hbar \omega \text{ }a^{\dagger }a+\frac{\hbar }{2}\omega _{a}\sigma _{z}+%
\frac{\hbar }{2}g\left( a^{\dagger }\sigma _{-}+a\sigma _{+}\right) , \label{hamilton}
\end{equation}%
where $a^{\dagger }$ and $a$ are the photon creation and annihilation operators respectively, and act on the photon
number state $|n\rangle $. The radiation-matter coupling constant $g$ is fixed by physical considerations such as the
cavity volume and the atomic dipole moment. The raising and lowering operators are defined by
\begin{eqnarray}
\sigma _{+} &\equiv &|e\rangle \langle f|,\text{\ }  \notag \\
\sigma _{-} &\equiv &|f\rangle \langle e|,\text{\ }  \label{prop5}
\end{eqnarray}%
and the $z$ Pauli operator by
\begin{equation}
\sigma _{z}\equiv |e\rangle \langle e|-|f\rangle \langle f|=\left[ \sigma _{+},\sigma _{-}\right] ,\label{prop6}
\end{equation}
and act on the atom states. Then the Hamiltonian, Eq. (\ref{hamilton}), is such that each photon creation is accompanied
by an atomic de-excitation, and each photon annihilation by an atomic excitation. For a given photon number value $n$,
Eq. (\ref{hamilton}) has the eigenvalues
\begin{equation}
E_{\pm }(n)=\hbar \omega (n+\frac{1}{2})\pm \frac{1}{2}\hbar \Omega _{n}(\delta ),  \label{auto}
\end{equation}%
where, for a specific detuning parameter $\delta \equiv \omega -\omega _{a} $,
\begin{equation}
\Omega _{n}(\delta )=\sqrt{\delta ^{2}+(n+1)g^2},  \label{Rabi}
\end{equation}%
is the Rabi frequency. The corresponding eigenvectors, called \textquotedblleft dressed states\textquotedblright , are
given by
\begin{eqnarray}
|n+\rangle  &=&\cos \theta _{n}|n,e\rangle +\sin \theta _{n}|n+1,f\rangle ,\text{\ }  \notag \\
|n-\rangle  &=&\cos \theta _{n}|n+1,f\rangle -\sin \theta _{n}|n,e\rangle, \text{\ }  \label{prop7}
\end{eqnarray}%
where
\begin{equation}
\tan (2\theta _{n})=\frac{g\sqrt{n+1}}{\delta }.  \label{teta}
\end{equation}%
Let us call $|\Psi (t)\rangle $ the wave function of the JCM system. Its dynamics is given by the time-dependent
Schr\"{o}dinger equation and we will procure general solutions of the form
\begin{equation}
|\Psi (t)\rangle =\sum_{n=0}^{\infty }\left( a_{n}^{+}e^{-\frac{iE_{+}(n)t}{\hbar }}|n+\rangle
+a_{n}^{-}e^{-\frac{iE_{-}(n)t}{\hbar }}|n-\rangle \right) ,  \label{Solu1}
\end{equation}%
where the coefficients $a_{n}^{+}$ and $a_{n}^{-}$ are fixed by the initial conditions. In this work we consider the case
of a separable atom-photon initial state. We assume that the initial state of the atom is arbitrary and that the initial
photon number follows some initial coherent distribution. However, we must be careful to eliminate the $\left\vert
{0}\right\rangle \left\vert{f}\right\rangle$ state in the initial condition because this state can not be built from the
dressed states, see Eq. (\ref{prop7}). More specifically, we take initial conditions of the form
\begin{equation}
|\Psi (0)\rangle  =\mathcal{N}\sum_{n=0}^{\infty }C_{n}\left\vert {n}\right\rangle\left( \left\vert {e}\right\rangle \cos {\frac{\gamma }{2}}+\left\vert {f}\right\rangle e^{i\varphi }\sin {\frac{%
\gamma }{2}}\right)-\mathcal{N}C_{0}\left\vert {0}\right\rangle\left\vert {f}\right\rangle e^{i\varphi }\sin
{\frac{\gamma }{2}} , \label{inipsi}
\end{equation}
where the two parameters $\gamma \in \left[ 0,\pi \right] $ and $\varphi \in \left[ 0,2\pi \right] $ define the initial
state of the atom characterized by a point on the generalized Bloch's sphere and $\mathcal{N}$ is a normalization
constant,
\begin{equation}
\mathcal{N}=\frac{1}{\sqrt{1-|C_{0}|^{2}{\sin ^{2}{\frac{\gamma }{2}}}}}. \label{norm}
\end{equation}
Here we take $\left\{|C_{n}|^{2}\right\}$ as the initial coherent photon distribution inside of the cavity without atoms.
Note that $\sum_{n=0}^{\infty}|C_{n}|^{2}=1$,  then $\overline{n} \equiv\sum_{n=0}^{\infty}n|C_{n}|^{2}$ is the initial
average number of photons. Using Eqs.(\ref{Solu1}),(\ref{inipsi}) and projecting on the base $\left\{ |n,e\rangle
,|n,f\rangle \right\} $, it is straightforward to obtain the connection between $\{a_{n}^{+},a_{n}^{-}\}$ and the initial
state of the atom and photon number,
\begin{eqnarray}
a_{n}^{+} &=&\mathcal{N}\left( C_{n}\cos \theta _{n}\cos {\frac{\gamma }{2}}+C_{n+1}\sin \theta _{n}e^{i\varphi }\sin
{\frac{\gamma }{2}}\right) ,
\notag \\
&&  \notag \\
a_{n}^{-} &=&\mathcal{N}\left( C_{n+1}\cos \theta _{n}e^{i\varphi }\sin {\frac{\gamma }{2}}-C_{n}\sin \theta _{n}\cos
{\frac{\gamma }{2}}\right) .
\notag \\
&&  \label{soo2}
\end{eqnarray}
For this system, as is the case with most closed quantum systems, the probability distribution does not converge in time.
However we can use a natural notion of convergence in the quantum case, if we define the \textquotedblleft limiting
distribution\textquotedblright\ as the asymptotic limit of the average of the probability distributions at time
\begin{equation}
P\equiv \lim_{t\mapsto \infty }\frac{1}{t}\int_{0}^{t}P(t)dt.  \label{def}
\end{equation}
This definition captures the amount of time the system spends in each average state, and moreover, it corresponds to the
natural concept of sampling from the system, since if one measures the system state at a random time chosen from the
interval $\left[ 0,t\right] $, the resulting distribution is exactly the average distribution. Applying the limit of the
average to the previous definitions, we calculate the probability to obtain $n$ photons
independently of the state of the atom, \emph{i.e.} $%
P(n)=\lim_{t\mapsto \infty }\frac{1}{t}\int_{0}^{t}\left\vert \left\langle {n}\right\vert \Psi (t)\rangle \right\vert
^{2}.$ After a straightforward calculation, we obtain the final photon distribution in thermal equilibrium with the atom
\begin{equation}
P(0)=|a_{0}^{+}|^{2}{\cos ^{2}{\theta _{0}}}+|a_{0}^{-}|^{2}{\sin ^{2}{\theta _{0}}},  \label{npos0}
\end{equation}
\begin{equation}
P(n) =|a_{n}^{+}|^{2}{\cos ^{2}{\theta _{n}}}+|a_{n}^{-}|^{2}{\sin ^{2}{\theta _{n}}}+|a_{n-1}^{+}|^{2}{\sin ^{2}{\theta
_{n-1}}}+|a_{n-1}^{-}|^{2}{\cos ^{2}{\theta _{n-1}}},  \label{nposs}
\end{equation}
for $n\geq 1$. This final photon distribution depends both of the initial state of the atom as well as of the initial
photon distribution, $\left\{|C_{n}|^{2}\right\}$. We can easily calculate the probability of finding the atom in its
excited (ground) state, independently of the number of photons in the system, \emph{i.e.}
\begin{equation}
P_{e} \equiv \lim_{t\mapsto \infty }\frac{1}{t}\int_{0}^{t}\sum_{n=0}^{\infty }\left\vert \left\langle {n,e}\right\vert
\Psi (t)\rangle \right\vert ^{2}dt=\sum\limits_{n=0}^{\infty }|a_{n}^{+}|^{2}{\cos ^{2}{\theta
_{n}}}+|a_{n}^{-}|^{2}{\sin ^{2}{\theta _{n}}}, \label{defa}
\end{equation}
\begin{equation}
P_{f} =1-P_{e}. \label{defb}
\end{equation}
Note that, using Eq.(\ref{soo2}),
\begin{equation}
|a_{n}^{+}|^{2} =\mathcal{N}^{2}\left( C_{n}^{2}\cos ^{2}{\theta _{n}}\cos ^{2}{\frac{\gamma
}{2}}+\frac{1}{2}C_{n}C_{n+1}\sin {2\theta _{n}}\sin {\gamma }\cos {\varphi }+ C_{n+1}^{2}\sin ^{2}{\theta _{n}}\sin
^{2}{\frac{\gamma }{2}}\right) ,  \label{ccu1}
\end{equation}
\begin{equation}
|a_{n}^{-}|^{2}=\mathcal{N}^{2}\left( C_{n}^{2}\sin ^{2}{{\theta _{n}\cos }^{2}{\frac{\gamma
}{2}}}-\frac{1}{2}C_{n}C_{n+1}\sin 2\theta _{n}\sin \gamma \cos \varphi+C_{n+1}^{2}\cos ^{2}{{\theta _{n}\sin
}^{2}{\frac{\gamma }{2}}}\right) ,  \label{cu2}
\end{equation}
then, the above probabilities are dependent both of the initial distribution of photons and of the initial state of the
atom .

In the asymptotic equilibrium, the average number of photons, $\left\langle n\right\rangle =\sum_{m=-1}^{\infty }nP(n)$,
also can be calculated. Using Eqs.(\ref{npos0}), (\ref{nposs}), (\ref{ccu1}), (\ref{cu2}), it is straightforward to show
that and the average number of photons in the equilibrium satisfies
\begin{equation}
\left\langle n\right\rangle = P_{f}+\left( {\overline{n}}-(1-e^{-\overline{n}}){%
\sin ^{2}{\frac{\gamma }{2}}}\right) /\left( 1-e^{-\overline{n}}\sin ^{2}{%
\frac{\gamma }{2}}\right).  \label{naverage}
\end{equation}
In the case $\overline{n}\gg1$, it is easy to show that the difference $\Delta n\equiv\left\langle
n\right\rangle-\overline{n}$ is
\begin{equation}
\Delta n\cong P_{f}-\sin ^{2}{\frac{\gamma }{2}}, \label{naverage2}
\end{equation}
so that, as it could be expected, the average number of photons does not change appreciably because of the presence of
the atom. The probability of finding the atom in its excited (ground) state in the resonant case is
\begin{equation}
P_{e}=P_{f}=\frac{1}{2},
\end{equation}
independent of the atomic initial conditions. In this case $\delta =0$ and $\theta _{n}=\pi /4$ for all $n$, see
Eqs.(\ref{teta}), (\ref{defa}) and (\ref{defb}).

In the case $\overline{n}\sim0$ we have the vacuum of the optical cavity and Eq.(\ref{naverage}) is reduced to
\begin{equation}
\left\langle n\right\rangle\cong P_{f}. \label{naverage3}
\end{equation}
In this case, we can study the thermal effects in the JCM using the already mentioned TFD formalism, according to which
there is a correspondence between the ensemble averages of statistical mechanics and the vacuum expectation values of
quantum field theory \cite{Thermo0}. This formalism was applied to the JCM in Refs. \cite{Thermo,Thermo1}.
Ref.\cite{Thermo1} obtains an explicit expression for the atomic state population as a function of temperature, namely
\begin{equation}
P_{e}=\frac{e^{\beta\hbar\omega/2}}{e^{\beta\hbar\omega/2}+e^{-\beta\hbar\omega/2}}, \label{defatfd}
\end{equation}
\begin{equation}
P_{f} =\frac{e^{-\beta\hbar\omega/2}}{e^{\beta\hbar\omega/2}+e^{-\beta\hbar\omega/2}}. \label{defetfd}
\end{equation}
where $\beta=(k_{B}T)^{-1}$, with $k_{B}$ the Boltzmann constant.
\section{Entanglement and temperature}
The unitary evolution of the JCM generates entanglement between the number of photons and the state of the atom. To
characterize this entanglement we start with the von Neumann entropy which is the quantum analogue of the Gibbs entropy
\begin{equation}
S_{N}(\rho )=-\mathrm{tr}(\rho \log \rho ).  \label{uno}
\end{equation}
where $\rho(t) =|\Psi (t)\rangle \langle \Psi (t)|$ is the density matrix of the quantum system. Due to the unitary
dynamics of the JCM the system remains in a pure state and this entropy vanishes. However, for these pure states the
entanglement between the state of the atom and the number of photons can be quantified by the associated von Neumann
entropy for the reduced density operator that defines the entropy of entanglement
\begin{equation}
S(\rho )=-\mathrm{tr}(\rho _{c}\log \rho _{c}),  \label{dos}
\end{equation}
where here we define $\rho_c$ as
\begin{equation}
\rho _{c}\equiv\lim_{t\mapsto\infty}\frac{1}{t}\int_{0}^{t}\sum_{n=0}^{\infty}\left\langle {n}\right|
\rho(t)\left|{n}\right\rangle ,  \label{dosb}
\end{equation}
and the partial trace is taken over the atom states. Using the wave function Eq. (\ref{Solu1}), together with Eq.
(\ref{dosb}),  we obtain the reduced density operator
\begin{equation}
\rho_{c} =\left(
\begin{array}{cc}
P_e & 0 \\
0 & P_f%
\end{array}%
\right) ,  \label{rho}
\end{equation}
Equation (\ref{rho}) is diagonal and we define the eigenvalues
\begin{equation}
\Lambda _{+}\equiv P_e,  \label{lam0}
\end{equation}
\begin{equation}
\Lambda _{-}\equiv P_f,  \label{lam1}
\end{equation}
For a given value of $\delta\neq0$, these eigenvalues depend on the initial condition of the atom through the angles
$\gamma$ and $\varphi$ and also on the initial photon distribution, see Eqs.(\ref{defa}) and (\ref{defb}).

In the resonant case, $\delta=0$, the system stays in the degenerate case $\Lambda_{+}=\Lambda_{-}=1/2$, and in this case
the entanglement entropy is maximum, $S(\rho)=1$. If we want a more complete description of this equilibrium in the
asymptotic limit, we need to know whether it is possible to associate a temperature to this entanglement entropy. In
order to answer this question, we follow references \cite{alejo2012,gustavo,renato2014}. According to the procedure
described in them, it is necessary to connect the eigenvalues of $\rho_c$ with an unknown associated Hamiltonian operator
$H_c$. To obtain this connection we use the quantum Brownian motion model of Ref.\cite{Kubo}. There one considers the
system associated with the atomic degrees of freedom, (atomic states characterized by the density matrix $\rho_c$) in
thermal contact (entanglement) with the bath system associated to the photons. The number of photons can be thought as an
external degree of freedom, in interaction with the internal degree of freedom, the atom state. This picture is
equivalent to the thermal contact between one system and its heat bath. In equilibrium
\begin{equation}
[H_c,\rho_{c}]=0,  \label{scho2}
\end{equation}
should be satisfied. As a consequence, in the asymptotic regime the density operator $\rho_{c}$ is an explicit function
of a time-independent Hamiltonian operator. If we note by $\{\left\vert\Phi_{\pm}\right\rangle\}$
the set of eigenfunctions of the density matrix, the operators $H_{c}$ and $%
\rho_{c}$ are both diagonal in this basis. Therefore the eigenvalues $%
\Lambda _{\pm}$ depend on the corresponding eigenvalues of $H_c$. We denote this set of eigenvalues by
$\{\epsilon_\pm\}$; they can be interpreted as the possible values of the entanglement energy. This interpretation agrees
with the fact that $\Lambda_{\pm}$ is the probability that the system is in the eigenstate
$\left\vert\Phi_{\pm}\right\rangle$. The precise dependence between $\Lambda _{\pm}$ and $\epsilon_\pm=\pm \epsilon$ is
determined by the type of ensemble. In our case, this equilibrium corresponds to a quantum canonical ensemble. Therefore
we propose that
\begin{equation}
\Lambda _{\pm}\equiv\frac{e^{\pm\beta\epsilon}}{e^{\beta\epsilon}+e^{-\beta%
\epsilon}},  \label{lam2}
\end{equation}
which defines the entanglement temperature $T\equiv1/(k_B\beta)$. Of course in Eq.(\ref{lam2}) only the ratio
$\epsilon/T$ is well defined, however we choose to introduce this temperature as this concept strengthens the idea of
asymptotic equilibrium between the photon distribution and the state of the atom. Note that while temperature makes sense
only in the mentioned equilibrium state, the entropy concept can be introduced without such a restriction.

The probability that a state chosen at random from the ensemble $\{\Phi_{+} $, $\Phi_{-}$\}, possesses an energy
$\epsilon$ is determined by the Boltzmann factor $e^{-\beta\epsilon}$. Let us call $\widetilde{\rho_{c}}$ the diagonal
expression of the density operator $\rho_{c}$, in the case treated in this paper $\widetilde{\rho_{c}}=\rho_{c}$, then
\begin{equation}
\widetilde{\rho_{c}}=\left(
\begin{array}{cc}
\Lambda _{+} & 0 \\
0 & \Lambda _{-}%
\end{array}%
\right) =\frac{1}{e^{\beta\epsilon}+e^{-\beta\epsilon}}\left(
\begin{array}{cc}
e^{\beta\epsilon} & 0 \\
0 & e^{-\beta\epsilon}%
\end{array}
\right) .  \label{rhoni}
\end{equation}
This operator is formally the same density operator that corresponds to an electron which possesses an intrinsic spin and
a magnetic moment in an external magnetic field \cite{pathria}. In general the Hilbert space of a quantum mechanical
model factors as a tensor product $\mathcal{H}_{sys}\otimes \mathcal{H}_{env}$ of the spaces describing the degrees of
freedom of the system and the environment. The evolution of the system is determined by the reduced density operator that
results from taking the trace over $\mathcal{H}_{env}$ to obtain $ \varrho_{sys}=\mathrm{tr}_{env}(\rho )$. The simple
models studied in Refs. \cite{Zurek1,Zurek2,Meyer} show how the correlations of a quantum system with other systems may
cause one of its observable to behave in a classical manner. In this sense the fact that the partial trace over the
photon number leads to a system effectively in thermal equilibrium, agrees with those previous results. Starting from
Eq.(\ref{rhoni}) it is possible to build the thermodynamics for the JCM entanglement, where the partition function of the
system is then given by
\begin{equation}
\mathcal{Z}=e^{\beta\epsilon}+e^{-\beta\epsilon}=2\cosh({\beta\epsilon}), \label{betaA}
\end{equation}
and the entanglement temperature is determined by
\begin{equation}
T=\frac{2\epsilon}{k_B\ln\left(\frac{\Lambda _{+}}{\Lambda _{-}}\right)}. \label{betae}
\end{equation}
When the system is in the resonance, $\Lambda _{\pm}=1/2$, it has maximum entropy and the entanglement temperature is
also maximum, $T=\infty$. It is important to point out that from the partition function Eq.(\ref{betaA}) it is possible
to develop all the entanglement thermodynamics, in other words, build the Hemholtz free energy, the internal energy and
the entropy. As it should be expected, the entropy obtained from the partition function agrees with the previous Shannon
expression, Eq.(\ref{ttres}). Using Eqs.(\ref{lam0}), (\ref{lam1}) and (\ref{lam2}), it is straightforward to show that
\begin{equation}
P_e=\frac{e^{\beta\epsilon}}{e^{\beta\epsilon}+e^{-\beta%
\epsilon}},  \label{pe}
\end{equation}
\begin{equation}
P_f=\frac{e^{-\beta\epsilon}}{e^{\beta\epsilon}+e^{-\beta%
\epsilon}}. \label{pf}
\end{equation}
\begin{figure}[b]
\begin{center}
\includegraphics[scale=0.3]{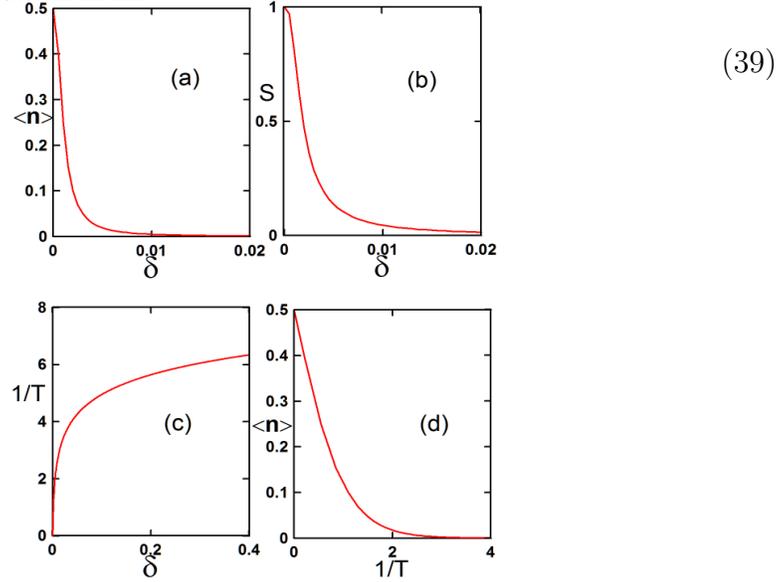}
\end{center}
\caption{ (a) Change in the average number of photons, (b) dimensionless entropy, and (c) inverse of the dimensionless
entanglement temperature, $1/T$, as a function of $\delta$. (d) Dependence of $\left\langle n\right\rangle$ on $1/T$. In
all these calculations we took $g=0.001$ and $\overline{n}=0.$ } \label{f1}
\end{figure}
It is interesting to make the connection between Eqs. (\ref{pe}) and (\ref{pf}) with the equivalent expressions obtained
within the TFD formalism, Eqs. (\ref{defatfd}) and (\ref{defetfd}). From a straightforward comparison of these equations,
we can determine explicitly the eigenvalues of $H_c$
\begin{equation}
\epsilon=\hbar\omega/2, \label{relation0}
\end{equation}
and we conclude that the thermal vacuum temperature as defined in TFD and the entanglement temperature are essentially
the same quantity, at least in the case of the JCM. The entanglement entropy, Eq.(\ref{dos}), can be expressed through
the eigenvalues of $\rho_{c}$ as
\begin{equation}
S(\rho)=-\Lambda_{+}\log_2 \Lambda_{+}-\Lambda_{-}\log_2 \Lambda_{-}. \label{ttres}
\end{equation}
\begin{figure}[t]
\begin{center}
\includegraphics[scale=0.3]{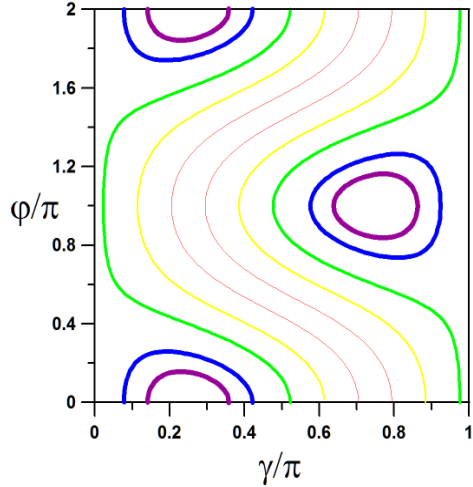}
\end{center}
\caption{(Color online) Five isothermal curves as  function of the initial conditions, the dimensionless angles
$\protect\gamma$ and $\protect\varphi$. Each curve is characterized by its thickness (and color).  The values of
$\beta\propto1/T$, from the thickest to the thinnest line are: 0.8 (purple), 0.7 (blue), 0.5 (green), 0.3 (yellow) and
0.1 (red). The isotherm corresponding to $\beta=0$ ($T=\infty$) is situated between the two red isotherms. Here $g=0.001,
\overline{n}=100, \delta=0.01$.} \label{f5}
\end{figure}
\begin{figure}[b]
\begin{center}
\includegraphics[scale=0.3]{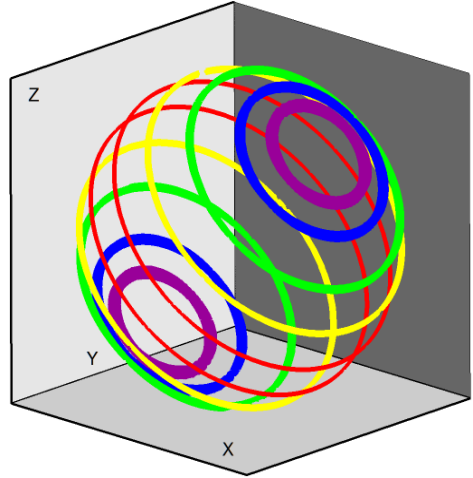}
\end{center}
\caption{(Color online) The isothermal curves of Fig.~\ref{f5} are shown on the Bloch sphere. The rotation axis of the
figure is in the direction $(1,0,1)$. The temperature of the entanglement between the atom and the photons is determined
by the atomic initial condition.} \label{f6}
\end{figure}
Our results, for $\overline{n}\sim0$, are presented graphically in Fig. \ref{f1}, which shows the behavior of
$\left\langle n\right\rangle$ as a function of the resonant parameter $\delta$. Figure \ref{f1}(a) shows that for any
value of $\delta$, $\left\langle n\right\rangle < 1/2$. Figure \ref{f1}(b) shows that the entanglement entropy has its
maximum value for the resonant case \emph{i.e.} $S = 1$ when $\delta=0$. The maximum value of the entropy (maximum
disorder) is achieved when the dimensionless entanglement temperature is $T=\infty$, see Fig. \ref{f1}(c). Under these
conditions the atom excitation probability behaves as a classical Markov process. The relation between $\left\langle
n\right\rangle$ and the inverse of the dimensionless entanglement temperature, shown in Fig. \ref{f1}(d), illustrates
that, for large values of the entanglement temperature, a proportionality exists between $\left\langle n\right\rangle$
and $1/T$. On the other hand for small values of $T$, $\left\langle n\right\rangle$ saturates to a constant value $0$.
These results are in accordance with Eq.(\ref{naverage}). For $\overline{n}\gg1$, figures \ref{f5} and \ref{f6} show the
isotherms for the entanglement temperature as a function of the atomic initial condition, Eq.(\ref{inipsi}). In Fig.
\ref{f5} the initial position is defined through the angles $\gamma$ and $\varphi$ and in Fig. \ref{f6} it is defined
through the position on the Bloch sphere.
\section{Conclusions}
We studied the unitary evolution of the JCM in a composite Hilbert space. In particular, we
considered the entanglement between a single two-level atom and a photon distribution characterized by a single
frequency.

The JCM probability distribution does not converge with time. However, the asymptotic limit of the time-average of the
probability distributions does converge. Thus the system establishes a stationary entanglement between the state of the
atom and the photon distribution that allows to develop a thermodynamic theory. The asymptotic reduced density operator
is used to introduce the entanglement thermodynamic functions in the canonical equilibrium. These thermodynamic functions
characterize the asymptotic entanglement. The JCM can be put into correspondence with a quantum walk on the line, where
the internal degree of freedom (the state of the atom) is entangled with the position degree of freedom (the photon
number).

The difference between the average number of photons, with and without photon-atom interaction, depends on the
entanglement temperature, and we show a connection between the entanglement temperature and the electromagnetic field
intensity in the optical cavity.

We remark that in the case studied in this paper the entanglement temperature between the photons and the atom coincides
with the temperature obtained within the TFD thermodynamics for the JCM. This implies that the entanglement temperature
is amenable to direct measurements.
\bigskip
We acknowledge the support from PEDECIBA (Programa de Desarrollo de las Ciencias B\'asicas) and ANII (Agencia Nacional de
Investigaci\'on e Innovaci\'on) (FCE-2-211-1-6281, Uruguay), and the CAPES-UdelaR collaboration program.  A.R.
acknowledges the stimulant comments of Victor Micenmacher.

\end{document}